# Segregation of Granular Mixtures in a Rotating Drum

by

Sanjay Puri[1,2] and Hisao Hayakawa[1]

[1] Graduate School of Human and Environmental Studies

Kyoto University, Sakyo-ku, Kyoto 606-01

JAPAN.

[2] School of Physical Sciences

Jawaharlal Nehru University

New Delhi – 110067

INDIA.

**Abstract**

We formulate a phenomenological model for the segregation of binary mixtures of rough and smooth granular materials in a rotating drum. Our model successfully replicates a range of experimental behaviours, e.g., rapid radial segregation; slow axial segregation; and nonuniform radial structure of axial bands. We present some analytical results and detailed numerical results for our model.

# 1 Introduction

The behaviour of granular materials has attracted much interest in the literature [1]. In large part, this is a consequence of the great technological relevance of such materials. At the physics level, granular matter throws up many challenging problems. For example, granular materials exhibit a range of complex phenomena such as convection [2], size segregation [3], pattern formation under vertical vibration [4], etc. In this paper, we will address one such problem which has been the subject of many experimental studies [5-14], viz., the dynamical segregation of a homogeneous mixture of two granular materials in a rotating drum.

There have also been a few numerical investigations of this problem. For example, in addition to their detailed experiments, Ueda et al. [13, 14] have also obtained preliminary numerical results on two different models, viz., (i) a "microscopic" cellular automaton (CA) model; and (ii) a coarse-grained phenomenological model for axial segregation, which is in the same universality class as the 1-dimensional Cahn-Hilliard (CH) equation for phase separation in binary mixtures. Furthermore, Levitan [15] has also presented numerical results for a phenomenological model of axial segregation. However, neither of the above studies has attempted to quantitatively characterize the asymptotic behaviour of axial segregation. More seriously, to the best of our knowledge, there is no comprehensive phenomenological model which captures the various experimental observations described below. In this paper, we formulate such a model and present results from analytical and numerical investigations of this model.

At the outset, we should stress an important point. There appears to be a general consensus [9, 13-16] that the difference in Coulombic friction between the two granular species plays an important role in the segregation process – though this has been disputed by Prigozhin and Kalman [17]. We shall follow the general consensus and consider the segregation of a mixture of rough (say, sand or S) and smooth (say, glass or G) materials. (Of course, the two materials could even have the same composition but merely differ in size and shape of the grains.)

Before we proceed, let us summarize the experimental and numerical observations in this context. These are as follows :

(a) There is an extremely rapid radial segregation [6, 7, 10, 13] with the smooth material (G) accumulating at the walls of the drum; and the rough material (S) accumulating in the central region. For example, this is clearly seen in the CA simulations of Ueda et al. (cf. Fig. 2(a) of Ref. [13]). Typically, the time-scale for this regime is a few rotation time-periods.

(b) The radially-segregated profile set up in (a) may become unstable on a longer time scale and is then followed by a slower axial segregation [9, 10, 13, 14], where the system phase-separates into alternating bands rich in S and G. These bands coarsen slowly in time, as depicted in Fig. 1 of Ref. [9]; Fig. 2 of Ref. [14]; and Fig. 2(b) of Ref. [13].

(c) A close inspection of the experimental pictures (e.g., Fig. 1 of Ref. [9]; or Fig. 2 of Ref. [14]) suggests that the bands in (b) are not uniform in the radial direction, being thicker (or thinner) at the centre of the drum.

We should clarify that every experiment does not necessarily show all the features mentioned in (a)-(c) above. Typically, some experiments do not see the rapid radial segregation regime, whereas other experiments may exhibit only radial segregation. Furthermore, the degree of radial non-uniformity of band size may differ from one experiment to another. In this paper, we present a simple phenomenological model which can account for all these possibilities.

This paper is organized as follows. In Sec. 2, we explain our phenomenological modelling and clarify the physical interpretation of parameters in our model. Section 2 also presents some analytical results for static solutions of our model. In Sec. 3 of this paper, we present results from numerical simulations of our model, which capture the various features described in (a)-(c) above. Finally, Sec. 4 ends this paper with a summary and discussion of our results and directions for future investigation.

## 2 Phenomenological Model for Segregation in a Rotating Drum

### 2.1 Phenomenological Model

Figure 1 is a schematic (in the radial cross-section) of a granular mixture (of G and S) in a rotating drum with radius $R$. This figure will serve as the basis of our subsequent discussion. The axis of the drum is horizontal and the drum is rotated counterclockwise about its axis with an angular velocity $\omega$. The surface of the rotated granular mixture is (in general) characterized by an $S$-shaped profile in the frequency regime relevant to the segregation problem. Typically, the flow of granular matter occurs in a thin laminar region at the surface [18, 9]. As a reference frame, we will use the inclined plane corresponding to the average surface profile of the homogeneous mixture when the drum is rotated. The coordinates $x$ and $y$ are measured in this frame in the radial and axial directions, respectively.

In the absence of any rotation, the granular mixture lies at an angle $\theta_0$ to the horizontal, determined as

$$\tan\theta_0 = \mu_G(1-\phi_S) + \mu_S\phi_S, \qquad (1)$$

where $\mu_G$ and $\mu_S$ are the (Coulombic) friction coefficients of glass and sand, respectively. In Eq. (1), $\phi_S$ refers to the number fraction of sand. For simplicity of presentation, we will focus on the case where the drum is half-filled with the granular mixture. However, our treatment below easily generalizes to the case of arbitrary filling fractions.

Our phenomenological model (following Ueda et al. [13, 14]) is based on a time-dependent order parameter field $\psi(\vec{r},t)$, which describes the relative concentrations of S and G in the 2-dimensional reference frame defined above. We use the convention that regions with $\psi > 0$ ($\psi < 0$) are rich in S (G), i.e., $\phi(\vec{r},t) = (1 + \psi(\vec{r},t))/2$, where $\phi(\vec{r},t)$ is the local number density of sand. At present, our modelling focuses on the laminar surface layer of the granular mixture and neglects potentially interesting effects in the bulk of the mixture. In future work, we intend to investigate the complete 3-dimensional problem.

The key ingredient of our phenomenological model is the intrinsic difference between the angles of repose for S and G, due to their different friction coefficients. We make the reasonable assumption that G, which is smooth, prefers to be located in regions with lesser angles of repose. Furthermore, the only relevant angle of repose is that along the radial direction. In the axial direction, fluctuations of the horizontal surface do not result in sufficient steepness to effect an axial flow of particles. Thus, we introduce the radial slope function $s(\vec{r},t)$, which measures the inclination of the surface to the horizontal at a point $\vec{r}$ in the reference frame. Then, there is a current of glass particles, which is driven by the gradient of the difference between the local slope $s(\vec{r},t)$ and the slope $s_G(\vec{r})(\equiv s_G(x))$ corresponding to the $S$-shaped profile for pure glass. (As any mixture of S and G has a higher friction coefficient than that for pure G, we expect that $s(\vec{r},t) \geq s_G(x)$. We will elaborate on this point at a later stage.) Naturally, the current of glass particles is from regions of higher slope difference to those of lower slope difference. By definition, the current of $\psi(\vec{r},t)$ flows in the opposite direction, viz.,

$$\vec{J}_{\rm sl} = M_0(1 - \psi(\vec{r},t)^2)\vec{\nabla}[s(\vec{r},t) - s_G(x)], \qquad (2)$$

where the subscript "sl" refers to current due to the "slope gradient". In Eq. (2), $M_0$ is the appropriate mobility factor which is proportional to the "temperature" $T$ of the granular mixture. The system evolves via stochastic S-G interchanges, with the overall composition being conserved, and this results in the order-parameter-dependence of the mobility through the factor $(1 - \psi^2)$. This factor goes to zero in regions which contain only pure S or G. In such regions, stochastic exchanges merely interchange identical particles and do not result in any evolution of the order parameter.

Before we proceed, two remarks are in order. Firstly, we have to carefully interpret the meaning of "temperature" in the case of granular mixtures. The usual thermodynamic temperature is not relevant in this context because of the relatively large sizes of grains. In the present context, "temperature" and stochastic exchanges arise due to granular collisions in the laminar layer. We will subsequently interpret "temperature" in terms of the angular velocity of the rotating drum. Secondly, our modelling does not include the effects of gravitational buoyancy due to density mismatches of the granular species. In most experiments on this problem, the densities of the two species are well-matched so the effect of gravitational buoyancy is negligible. In any case, it is straightforward to include a current due to gravity in our modelling.

In addition to the above current, there is also a random mixing of particles due to diffusion in the laminar layer. As with our previous parameters, diffusion also has a dynamical origin and results from granular collisions in the flowing laminar layer. In our modelling, segregation results in certain parameter regimes from the competition between diffusive mixing and the intrinsic tendency for S and G to incline at different slopes.

Finally, we penalize interfaces between S-rich and G-rich domains. In general, the grains of S and G have different shapes and sizes, and stresses result due to mismatches at interfacial regions between domains rich in the dissimilar materials. The stresses involved are long-ranged, in general, but we confine ourselves to the simplest case of a "square-gradient" interfacial penalty. As in the familiar case of the CH equation, this leads to a current contribution

$$\vec{J}_{\text{ip}} = K(1 - \psi(\vec{r},t)^2)\vec{\nabla}[\nabla^2 \psi(\vec{r},t)], \tag{3}$$

where the subscript "ip" refers to current due to the interfacial penalty; and $K$ is a phenomenological constant. In our phenomenological modelling, this term has the effect of stabilizing and smoothing out domain interfaces. This is in accordance with our physical expectation that there will be a smooth transient regime between S-rich and G-rich domains due to avalanches of particles. We should further stress that our model has divergent solutions for a range of parameter values in the limit $K = 0$.

Thus, the temporal evolution of the order parameter is modelled as follows :

$$\frac{\partial \psi(\vec{r},t)}{\partial t} = -\vec{\nabla} \cdot (\vec{J}_{\text{sl}} + \vec{J}_{\text{ip}}) + D\nabla^2 \psi(\vec{r},t), \qquad (4)$$

where $D$ is the phenomenological diffusion constant, which is proportional to the temperature $T$. Subsequently, we will discuss the dependence of $M_0$ and $D$ on the rotation frequency $\omega$. Replacing the expressions for currents from Eqs. (2)-(3) in Eq. (4), we obtain

$$\begin{aligned}\frac{\partial \psi(\vec{r},t)}{\partial t} &= -\vec{\nabla} \cdot \left\{ (1 - \psi(\vec{r},t)^2)\vec{\nabla}\left[M_0(s(\vec{r},t) - s_G(x)) + K\nabla^2 \psi(\vec{r},t)\right] \right\} \\ &\quad + D\nabla^2 \psi(\vec{r},t). \end{aligned} \qquad (5)$$

The final ingredient in our modelling is the behaviour of the radial slope function $s(\vec{r},t)$. The local shape of the surface profile is established as the result of a balance between flow in the laminar layer and accretion (depletion) of particles from (to) the rotating bulk. The basic mechanism for this has been discussed in the context of a one-component mixture by Zik et al. [9]. We will make the assumption that the shape of the surface profile is in local equilibrium at all times. This is reasonable as segregation results via particle diffusion, which occurs on much slower time-scales than the flows which establish the surface profile. In the next subsection, we employ simple arguments to obtain the general shape of the surface profile.

## 2.2 Instantaneous Shape of the Surface Profile

Let us first consider the case of a one-component granular material in a rotating drum. As we have stated earlier, we focus on the case of a half-filled drum though the arguments below easily generalize to the case of arbitrary filling fraction. From experiments, we know that the surface is characterized by an S-shaped profile, as depicted in Fig. 1. The steady-state profile for a one-component material will be independent of the axial coordinate. If we restrict ourselves to the case of a symmetric S-shaped profile in $x \to -x$, we have a general expression for the local slope $s(x) = \tan[\theta(x)]$ as follows :

$$s(x) = \mu + f(\omega)(R^2 - x^2) + g(\omega)(R^2 - x^2)^2 + \text{higher order terms}, \quad (6)$$

where $\mu$ is the Coulombic friction coefficient of the material; and $f(\omega), g(\omega) \to 0$ as $\omega \to 0$. Notice that the expansion in Eq. (6) sets $s(\pm R) = \mu$ because there is no radial current in the laminar layer at $x = \pm R$. Therefore, we expect the granular material to be in static repose at the drum edges. Furthermore, we expect $s(x) \geq \mu$ for all $x$ on rotation of the drum. The simple expansion in Eq. (6) does not account for asymmetries in the profile resulting from the fact that different friction coefficients are applicable at the top and bottom. The appropriate friction coefficients at the top $(x = R)$ and bottom $(x = -R)$ are the static $(\mu_s)$ and dynamic $(\mu_d < \mu_s)$ coefficients, respectively. This is easily incorporated in generalized versions of the model presented here.

We can also consider clockwise rotations of the drum with angular frequency $-\omega$. From physical considerations, it is clear that $\theta(-\omega, x) = \pi - \theta(\omega, x)$ or $s(-\omega, x) = -s(\omega, x)$. This does not impose any further constraints on the form of $s(x)$ in Eq. (6) but merely imposes an overall factor of $\text{sgn}(\omega)$ as $s(\omega, x) = \text{sgn}(\omega) s(|\omega|, x)$. Without any loss of generality, we will only consider counterclockwise rotations of the drum.

Finally, we Taylor-expand $f(\omega), g(\omega)$, etc. in Eq. (6) and retain only the first nontrivial term in the expansion to obtain

$$s(x) \simeq \mu + f_1 \omega (R^2 - x^2) + \text{higher order terms}, \tag{7}$$

where $f_1 = f'(0) > 0$. In Appendix A, we present arguments due to Zik et al. [9], who use a "microscopic" approach to determine the shape of the S-shaped profile. This facilitates a physical identification of parameters in the above expansion. We also extend the arguments of Zik et al. [9] to the case of a two-component granular material in a rotating drum. Of course, we should stress that their assumptions are of limited validity, as we clarify in Appendix A. Nevertheless, their approach serves as a useful guide to obtain a phenomenological expression for the S-shaped profile of rotated granular matter – both in the one-component and two-component cases.

Next, we consider the generalization of Eq. (7) to the case of a two-component granular material. Under the assumption of local equilibrium, the expression in Eq. (7) is applicable to the two-component case also but with appropriate generalizations as $\mu \to \mu(\psi)$ and $f_1 \to f_1(\psi)$. The simplest functional form for the friction coefficient of the mixture is

$$\mu = \frac{\mu_S + \mu_G}{2} + \frac{\mu_S - \mu_G}{2} \psi \equiv \mu_+ + \mu_- \psi. \tag{8}$$

The quantity $f_1(\psi)$ can be Taylor-expanded in the weak-segregation limit (where $\psi$ is small) as

$$f_1(\psi) = a_0 + a_1 \psi + a_2 \psi^2 + a_3 \psi^3 + \text{higher order terms}. \tag{9}$$

Replacing Eqs. (8)-(9) in Eq. (7), we obtain a general expression for the local slope function as

$$\begin{aligned} s(\vec{r}, t) &= \mu_+ + \mu_- \psi(\vec{r}, t) + \\ &\quad [a_0 + a_1 \psi(\vec{r}, t) + a_2 \psi(\vec{r}, t)^2 + a_3 \psi(\vec{r}, t)^3] \omega (R^2 - x^2) + \\ &\quad \text{higher order terms}. \end{aligned} \tag{10}$$

The appropriate expression for pure glass, $s_G(x)$, is obtained by setting $\psi = -1$ in Eq. (10) to obtain

$$s_G(x) = \mu_+ - \mu_- + (a_0 - a_1 + a_2 - a_3)\omega(R^2 - x^2) + \text{higher order terms.} \quad (11)$$

Therefore, we have

$$\begin{aligned}s(\vec{r},t) - s_G(x) &= \mu_- + (a_1 - a_2 + a_3)\omega(R^2 - x^2) + \\ &\quad [\mu_- + a_1\omega(R^2 - x^2)]\psi(\vec{r},t) + a_2\omega(R^2 - x^2)\psi(\vec{r},t)^2 + \\ &\quad a_3\omega(R^2 - x^2)\psi(\vec{r},t)^3 + \text{higher order terms.} \quad (12)\end{aligned}$$

For a reasonable physical identification of the various expansion parameters, we refer the interested reader to Appendix A. It should also be kept in mind that we expect $s(\vec{r},t) - s_G(x) \geq 0$ on physical grounds.

## 2.3 Dynamical Model for Segregation

Replacing the expansion for $(s(\vec{r},t) - s_G(x))$ from Eq. (12) in Eq. (5), we obtain the following dynamical equation for the order parameter :

$$\begin{aligned}\frac{\partial \psi(\vec{r},t)}{\partial t} &= -\vec{\nabla} \cdot \left\{(1 - \psi(\vec{r},t)^2)\vec{\nabla}\left[a(R^2 - x^2) + [b + c(R^2 - x^2)]\psi(\vec{r},t)\right.\right.\\ &\quad \left.\left. + d(R^2 - x^2)\psi(\vec{r},t)^2 + e(R^2 - x^2)\psi(\vec{r},t)^3 + K\nabla^2\psi(\vec{r},t)\right]\right\} \\ &\quad + D\nabla^2\psi(\vec{r},t), \quad (13)\end{aligned}$$

where we have dropped the higher order terms in the expansion for the local slope. The parameters in Eq. (13) are identified as

$$\begin{aligned}a &= M_0(a_1 - a_2 + a_3)\omega, \\ b &= M_0\mu_-, \\ c &= M_0 a_1 \omega, \\ d &= M_0 a_2 \omega, \\ e &= M_0 a_3 \omega. \quad (14)\end{aligned}$$

At this stage, we should stress that the phenomenological model obtained above is expected to be of greater validity than the expansions invoked to arrive at the model. This is based on our experience with far-from-equilibrium systems, where phenomenological coarse-grained models are well-known to be of greater applicability than their "derivations" would suggest. It is in this spirit that we shall proceed. In the ultimate analysis, our model will only be justified in terms of its ability to replicate the phenomena we set out to describe.

For subsequent analysis, it is convenient to formulate Eq. (13) in a form equivalent to that of the CH equation with an order-parameter-dependent mobility [21]. Towards this end, we make the identification

$$\nabla^2 \psi(\vec{r},t) = \vec{\nabla} \cdot \left\{ (1 - \psi(\vec{r},t)^2) \vec{\nabla} \left[ \tanh^{-1}[\psi(\vec{r},t)] \right] \right\}. \tag{15}$$

Then, we can rewrite Eq. (13) as

$$\begin{aligned}
\frac{\partial \psi(\vec{r},t)}{\partial t} &= -\vec{\nabla} \cdot \left\{ (1 - \psi(\vec{r},t)^2) \vec{\nabla} \Big[ a(R^2 - x^2) + [b + c(R^2 - x^2)] \psi(\vec{r},t) \right. \\
&\quad + d(R^2 - x^2) \psi(\vec{r},t)^2 + e(R^2 - x^2) \psi(\vec{r},t)^3 - D \tanh^{-1}[\psi(\vec{r},t)] \\
&\quad \left. + K \nabla^2 \psi(\vec{r},t) \Big] \right\} \\
&\equiv -\vec{\nabla} \cdot \left\{ (1 - \psi(\vec{r},t)^2) \vec{J}(\vec{r},t) \right\}. \tag{16}
\end{aligned}$$

Now, the basic structure of our model is clear. In the absence of rotation, we have $M_0, D, K = 0$ and the granular mixture does not evolve in any fashion, as we would expect. The rotation of the drum gives rise to granular motion and, thereby, the possibility of segregation. Thus, when the drum is rotated, there is an $x$-dependent current, which rapidly drives G to the drum walls and S to the drum centre (when $a > 0$). This corresponds to the regime referred to as (a) in our introductory section. The rotation also enhances the drive to segregate along the axial direction (when $c > 0$), with the maximum enhancement being at the centre of the drum. There is a competition between the radial current and the axial instability and this may result in axial segregation, as we will discuss shortly. The axial segregation corresponds to regime (b) in our introductory section. Finally, if axial segregation does occur, the drive for axial segregation is stronger at the centre of the drum, giving rise to the radially non-uniform shape of the axial bands, discussed in point (c) of our introduction.

In various limits, the general model in Eq. (16) also models experimental situations where one of the features mentioned in (a)-(c) of the introduction is absent. For example, if we set $a = 0$ in (16), there is no radial segregation. Alternatively, if we set $c = 0$ in (16), there is no radial non-uniformity in the axial bands. Furthermore, the terms proportional to $\psi(\vec{r}, t)^2$ in the current-determining term of Eq. (16) also account for segregation asymmetries.

Finally, we should point out that an experimentally relevant generalization of our model would incorporate the asymmetry of the $S$-shaped profile via an additional small term (relative to the $x^2$-term) linear in $x$ in the current-determining expression in (16).

At present, we wish to work with the simplest model which captures the broad experimental features discussed in the introduction. Therefore, we simplify the model in Eq. (16) by setting $d = e = 0$ (or $a_2 = a_3 = 0$), which also fixes $c = a$ (cf. Eq. (14)). It is not our thesis that this simplified model is in the same dynamical universality class as the complete model. Nevertheless, as we will see later, this simplified model already exhibits many experimentally relevant features. In subsequent work, we will undertake a more exhaustive study of the dynamics of our model in various regions of parameter space.

## 2.4 Interpretation of Parameters and Dimensionless Rescaling

It is relevant to ask about the meaning of various parameters in our model. In particular, we are interested in the interpretation of "temperature" $T$, which gives rise to stochastic motion of the granular particles, thereby enabling segregation. The usual thermodynamic temperature is not relevant here as the grains are too large to be moved about by thermal agitation. In the present context, motion of grains is the result of random collisions in the flowing laminar layer, which is (in turn) a consequence of drum rotation. Thus, we will interpret $T$ in terms of the average fluctuation of granular kinetic energy in the laminar layer – analogous to the definition of microscopic temperature in terms of the kinetic energy of atoms of a gas.

We use de Gennes' argument (reproduced in Appendix A) to compute the average velocity and the mean-square velocity of granular particles in the laminar layer. In Appendix A, we have obtained the radial dependence of the granular velocity. A simple calculation based on this yields

$$\begin{aligned} \langle v \rangle &= \frac{1}{3}\frac{\omega R^2}{h_0}, \\ \langle v^2 \rangle &= \frac{2}{15}\frac{\omega^2 R^4}{h_0^2}, \end{aligned} \quad (17)$$

where $h_0$ is the width of the laminar layer (assumed constant). The mean-square fluctuation in the velocity is $\sigma^2 = \langle v^2 \rangle - \langle v \rangle^2 = \omega^2 R^4/(45 h_0^2)$. Thus, we define the granular temperature in a rotating drum as $T = m\omega^2 R^4/(45 h_0^2)$, where $m$ is the mass of a granular particle. The corresponding self-diffusion constant is obtained by using the expression for a hard-sphere fluid [22] as follows :

$$D \propto \frac{m\omega R^2}{\rho d^2 h_0}, \quad (18)$$

where $\rho$ is the fluid density; and $d$ is the typical grain size. We have deliberately omitted prefactors as this expression is only reasonable for obtaining dimensional dependences. After all, the granular particles are not really hard spheres. Furthermore, the granular density is too high for the above expression to be numerically accurate.

Finally, the dimensional dependence of the mobility is obtained from Eq. (2) as $[M_0] = \ell_0^3 \ell^{-1} \tau^{-1}$, where $\ell_0$ is a coarse-graining length; and $\ell$ and $\tau$ are the scales of length and time, respectively. Thus, $M_0$ is also proportional to the rotation frequency as the average granular velocity sets the time-scale of particle collision and interchange.

We now rescale our model of Eq. (16) (with $d = e = 0$) into dimensionless units. Notice that the dimensional properties of various parameters are as follows :

$$[K] = \ell^4 \tau^{-1},$$
$$[D] = \ell^2 \tau^{-1}. \tag{19}$$

We use these as the basis of the following rescaling to dimensionless units :

$$\vec{r} = \left(\frac{K}{D}\right)^{1/2} \vec{r}',$$
$$t = \frac{K}{D^2} t',$$
$$R = \left(\frac{K}{D}\right)^{1/2} R',$$
$$a = \frac{D^2}{K} a',$$
$$b = D b'. \tag{20}$$

This yields the following dimensionless equation (dropping primes) :

$$\frac{\partial \psi(\vec{r},t)}{\partial t} = -\vec{\nabla} \cdot \left\{ (1 - \psi(\vec{r},t)^2) \vec{\nabla} \left[ a(R^2 - x^2) + [b + a(R^2 - x^2)] \psi(\vec{r},t) - \tanh^{-1}[\psi(\vec{r},t)] + \nabla^2 \psi(\vec{r},t) \right] \right\}. \tag{21}$$

If we use Eq. (14) and the dimensional dependences discussed above, we see that $b$ is independent of $\omega$, and $a \propto \omega$ in Eq. (21).

As we are interested in a system of finite extent in the radial direction, we must supplement Eq. (21) with appropriate boundary conditions in the radial direction. We impose flat boundary conditions on the order parameter as the first condition, i.e.,

$$\left.\frac{\partial \psi(\vec{r},t)}{\partial x}\right|_{x=\pm R} = 0. \tag{22}$$

Furthermore, we impose the physical condition that there should be no radial current at the drum boundary, viz.,

$$\left.\frac{\partial}{\partial x}\left[a(R^2-x^2)+[b+a(R^2-x^2)]\psi(\vec{r},t)-\tanh^{-1}[\psi(\vec{r},t)]+\nabla^2\psi(\vec{r},t)\right]\right|_{x=\pm R} = 0. \tag{23}$$

In subsequent discussions, we will focus on Eqs. (21)-(23), which are characterized by the dimensionless parameters $a$ and $b$; and the drum radius $R$.

## 2.5 Static Radial Solutions and Linear Stability Analysis

Let us consider the linear stability properties of the above model. First, we examine small fluctuations about an initially homogeneous mixture of equal quantities of S and G, viz., $\psi(\vec{r},t) = 0 + \delta\psi(\vec{r},t)$. Linearizing Eqs. (21)-(23) in the fluctuation field, we obtain the following system of equations :

$$\frac{\partial}{\partial t}[\delta\psi(\vec{r},t)] = 2a - \nabla^2\left[b + a(R^2 - x^2) - 1 + \nabla^2\right]\delta\psi(\vec{r},t), \tag{24}$$

$$\left.\frac{\partial}{\partial x}[\delta\psi(\vec{r},t)]\right|_{x=\pm R} = 0, \tag{25}$$

$$\left.\frac{\partial}{\partial x}\left[b + a(R^2 - x^2) - 1 + \nabla^2\right]\delta\psi(\vec{r},t)\right|_{x=\pm R} = \pm 2aR. \tag{26}$$

The state $\psi = 0$ does not constitute a static solution of Eqs. (21)-(23). Therefore, currents are set up which rapidly drive the solution to the static radial profile $\psi_s(x)$, which is obtained as the zero-current solution of Eqs. (21)-(23) as follows :

$$\frac{d^2\psi_s(x)}{dx^2} - \tanh^{-1}[\psi_s(x)] + [b + a(R^2 - x^2)]\psi_s(x) + a(R^2 - x^2) = C, \tag{27}$$

$$\left.\frac{d\psi_s(x)}{dx}\right|_{x=\pm R} = 0. \tag{28}$$

The integration constant $C$ is determined by the constraint that $\int_{-R}^{R} dx \psi_s(x) = 0$ for a mixture with equal amounts of S and G.

The solutions of Eq. (27) can be categorized in terms of trajectories of the non-autonomous dynamical system :

$$\frac{d\psi_s(x)}{dx} = y(x),$$
$$\frac{dy(x)}{dx} = C - a(R^2 - x^2) - [b + a(R^2 - x^2)]\psi_s(x) + \tanh^{-1}[\psi_s(x)], \tag{29}$$

where $x$ is the "time" variable. For a particular value of $C$, the relevant trajectory goes from $y = 0$ at $x = -R$ to $y = 0$ at $x = R$. As mentioned earlier, the appropriate value of $C$ is determined from the constraint on overall composition. Unfortunately, because of the nonlinearity and explicit $x$-dependence in Eq. (27), the static radial solution has to be determined numerically, in general.

In the next section, we will present numerical results for the case when $b < 1$ and $a \sim O(0.05)$. This is an interesting range of parameter values for our model because the case with $b < 1$ corresponds to the experimentally relevant situation where a critical value of $a$ (or rotation frequency) is needed to initiate segregation. At the same time, large values of $a(> 0.1)$ in our model correspond to a situation in which there is only radial segregation because the static radial profile cannot be destabilized by axial fluctuations. For this range of parameter values, we can obtain a reasonable approximation for the static radial solution by a perturbative expansion in the parameter

a. Recall that the static solution for $a = 0$ is simply $\psi_s(x) = 0$. We use this solution as the basis for an expansion :

$$\psi_s(x) = \sum_{n=1}^{\infty} a^n \phi_n(x),$$
$$C = \sum_{n=1}^{\infty} a^n C_n. \tag{30}$$

It is relatively easy to obtain the solution $\phi_1(x)$ upto $O(a)$, so we demonstrate this explicitly. Replacing this expansion back in Eqs. (27)-(28), we obtain the system of equations (correct to $O(a)$) :

$$\frac{d^2\phi_1(x)}{dx^2} - (1-b)\phi_1(x) = C_1 - (R^2 - x^2), \tag{31}$$

$$\left.\frac{d\phi_1(x)}{dx}\right|_{x=\pm R} = 0. \tag{32}$$

The homogeneous solutions of Eq. (31) are $\exp(\pm\beta x)$, where $\beta = \sqrt{(1-b)}$. We can use these solutions to obtain the inhomogeneous solution of Eq. (31) by the usual method of variation of parameters [23]. The general solution of Eq. (31) is as follows :

$$\phi_1(x) = d_1 e^{\beta x} + d_2 e^{-\beta x} - \frac{1}{\beta^2}(C_1 - R^2 + x^2) - \frac{2}{\beta^4}, \tag{33}$$

where $d_1$ and $d_2$ are arbitrary constants, which are determined from the boundary conditions in (32). Imposing the boundary conditions and fixing $C_1$ to satisfy the composition constraint, we obtain the $O(a)$ solution as

$$\phi_1(x) = \frac{2R}{\beta^3 \sinh(\beta R)} \cosh(\beta x) - \frac{1}{\beta^2}\left(x^2 - \frac{R^2}{3}\right) - \frac{2}{\beta^4}. \tag{34}$$

We have also obtained the solution correct to $O(a^2)$. This requires considerably more algebra and we merely quote the final result :

$$\phi_2(x) = e_1 e^{\beta x} + e_2 e^{-\beta x} + \phi_{2,p}(x), \tag{35}$$

where the particular solution is

$$\phi_{2,p}(x) = C_2' - \frac{2}{\beta^4}\left(\frac{2R^2}{3} - \frac{7}{\beta^2}\right)x^2 + \frac{1}{\beta^4}x^4 +$$
$$\frac{R}{\beta^4 \sinh(\beta R)}\left[\frac{1}{2\beta}\left(R^2 - \frac{1}{2\beta^2}\right)\cosh(\beta x) - \left(R^2 - \frac{1}{2\beta^2}\right)x\sinh(\beta x) -\right.$$
$$\left.\frac{1}{2\beta}x^2\cosh(\beta x) + \frac{1}{3}x^3\sinh(\beta x)\right]. \tag{36}$$

The arbitrary constants $e_1$ and $e_2$ are obtained as

$$e_1 = e_2 = -\frac{1}{2\beta\sinh(\beta R)}\frac{d\phi_{2,p}(x)}{dx}\bigg|_{x=R}. \tag{37}$$

Finally, the constant $C_2'$ is obtained from the constraint $\int_{-R}^{R} dx\,\phi_2(x) = 0$ as

$$C_2' = \frac{1}{\beta^2 R}\frac{d\phi_{2,p}(x)}{dx}\bigg|_{x=R} + \frac{11}{45}\frac{R^4}{\beta^4} - \frac{14}{3}\frac{R^2}{\beta^6} + \frac{15}{4}\frac{1}{\beta^8} +$$
$$\frac{1}{2\beta^5}\left(\frac{4}{3}R^2 - \frac{7}{\beta^2}\right)R\coth(\beta R). \tag{38}$$

In the next section, we will compare our numerical results with the approximate radial solution $\psi_s(x) = a\phi_1(x) + a^2\phi_2(x)$, with $\phi_1(x)$ and $\phi_2(x)$ given by Eqs. (34) and (35)-(38), respectively.

Finally, we consider the linear stability properties of the static radial solution $\psi_s(x)$. As usual, we consider small fluctuations around the static solution as $\psi(\vec{r}, t) = \psi_s(x) + \delta\psi(\vec{r}, t)$. Linearizing Eqs. (21)-(23) in these fluctuations, we obtain the following system of equations :

$$\frac{\partial}{\partial t}[\delta\psi(\vec{r},t)] = -\vec{\nabla}\cdot\left\{(1-\psi_s(x)^2)\vec{\nabla}\left[b + a(R^2 - x^2) -\right.\right.$$
$$\left.\left.\frac{1}{1-\psi_s(x)^2} + \nabla^2\right]\delta\psi(\vec{r},t)\right\}, \tag{39}$$

$$\frac{\partial}{\partial x}[\delta\psi(\vec{r},t)]\bigg|_{x=\pm R} = 0, \tag{40}$$

$$\frac{\partial}{\partial x}\left[b + a(R^2 - x^2) - \frac{1}{1-\psi_s(x)^2} + \nabla^2\right]\delta\psi(\vec{r},t)\bigg|_{x=\pm R} = 0. \tag{41}$$

Unfortunately, Eqs. (39)-(41) are not analytically tractable. We should stress that the primary complication in the analysis of this system of equations is the finite boundary in the radial direction, which reduces the stability analysis to the spectral analysis of a fourth-order differential operator on a finite domain. However, for certain parameter values, we expect the static solution to be destabilized by axial fluctuations. A necessary (though not sufficient) condition for this is that the quantity $b+a(R^2-x^2)-1/(1-\psi_s(x)^2) > 0$ for some value of $x \in [-R, R]$. Once this condition is satisfied, axial instabilities arise but these are opposed by radial currents. We are presently examining the stability problem through numerical analysis and will present detailed results on this at a later stage. In the present exposition, we confine ourselves to showing representative numerical results for parameter regimes which exhibit both radial and axial segregation.

## 3  Numerical Results

In this section, we present typical numerical results from simulations of our phenomenological model in Eqs. (21)-(23). The primary purpose of this section is to demonstrate that our model can replicate the experimental features discussed in (a)-(c) of the introductory section.

We will first focus on the dynamics of radial segregation. Thus, we consider the 1-dimensional version of Eqs. (21)-(23) obtained by neglecting the axial ($y$-) dependence. We implemented a simple Euler-discretized version of this model on a 1-dimensional line specified by $x \in [-R, R]$, with $R = 4$ dimensionless units. The discretization mesh sizes were $\Delta x = 0.1$ (in space) and $\Delta t = 0.00005$ (in time). The parameter values for our simulation were $b = 0.9$ and $a = 0.01, 0.03, 0.05$ and $0.1$. Recall that $b = 0.9$ corresponds to a situation in which there is no segregation when $a = 0$ and segregation is driven by the enhancement of the effective diffusion coefficient due to rotation.

The initial conditions for our simulations consisted of an order parameter field which was randomly and uniformly distributed (with amplitude 0.05) about a background value of 0. This mimics a homogeneous mixture of equal amounts of S and G with small local fluctuations. We have also obtained numerical results for mixtures with unequal amounts of S and G. For the sake of brevity, we will not present these results here.

Fig. 2 demonstrates the rapid radial segregation in our model for a homogeneous initial condition. The parameter values were $b = 0.9$ and $a = 0.03$. Fig. 2 plots the evolution of the radial profile for dimensionless times $t = 0.5, 1, 2, 4$ and 10. The profile equilibrates by $t = 10$ and the static solution is seen to be in good agreement with the analytic approximation obtained in subsection 2.5 (denoted by a solid line in Fig. 2), except in the central region. Recall that regions with $\psi > 0$ are rich in the rough material (S) and those with $\psi < 0$ are rich in the smooth material (G). For these parameter values, the central region consists of approximately 70 % S and 30 % G; whereas the composition in the wings is approximately 30 % S and 70 % G.

At present, we are not in a position to directly compare our numerical results with those from experiments as this would require a determination of relative time-scales in experiments and our simulations. However, if we consider that radial segregation in experiments occurs on a time-scale of (say) 2-3 rotations, this would fix our dimensionless unit of time (in Fig. 2) as corresponding to approximately 0.2-0.3 rotations.

Fig. 3 shows the static radial profiles obtained numerically for $b = 0.9$ and $a = 0.01, 0.03, 0.05$ and $0.1$. We also superpose the analytic approximation obtained in subsection 2.5 (denoted as a solid line) for the parameter values $a = 0.01, 0.03, 0.05$. For $a > 0.06$, our analytic approximation does not work well because it does not account for saturative nonlinearities, which are crucial in determining the solution for (say) $a = 0.1$. Fig. 3 does not show the analytic approximation for $a = 0.1$ as its amplitude overshoots that of the numerical profile – both in the central region and the wings.

We next consider results from a simulation of Eqs. (21)-(23) in the complete 2-dimensional case. Again, we implemented a simple Euler discretization of Eqs. (21)-(23) (with isotropized Laplacians) on a lattice of size $[-R, R] \times [0, L]$. As in the previous case, we set the drum radius $R = 4$. The axial length for the evolution pictures we show subsequently was $L = 128$. The discretization mesh sizes in this case were $\Delta x = 0.5$ and $\Delta t = 0.01$; and the parameter values were $b = 0.9$ and $a = 0.03$. In the radial direction, the boundary conditions were fixed by Eqs. (22)-(23). In the axial direction, we set periodic boundary conditions.

Fig. 4 depicts the temporal evolution of a homogeneous initial condition in the 2-dimensional simulation. There is a rapid radial segregation, which results in an $S$-rich central region and $G$-rich wings. As we have seen earlier, the time-scale for radial segregation is $t \sim 10$ for these parameter values. The first frame in Fig. 4 shows the radially segregated state at $t = 100$. At approximately $t = 1000$ (200-300 drum rotations), the radially-segregated state is destabilized by axial fluctuations and breaks into axial bands. In Fig. 4, this is depicted in frames 2-6, corresponding to dimensionless times $t = 1100, 1200, 1300, 1400$ and $1500$. These bands coarsen slowly in time, as shown in frames 7-9 at times $t = 1600, 10000$ and $20000$, respectively. The parameter values used here correspond to a situation in which the segregation drive is enhanced in the central region and this leads to a radial non-uniformity in the shape of the bands (cf. Fig. 2 of Ref. [14]). As a matter of fact, for these parameter values, there is no intrinsic drive towards segregation in the wings. Thus, segregation at the edges is dragged by segregation in the central region.

In Figs. 5(a)-(c), we show the axial variation of the order parameter $\psi(x, y, t)$ vs. $y$ for different values of the radial coordinate $x$. Fig. 5(a) corresponds to the case $x = 0$ (i.e., the centre of the drum) and plots $\psi(0, y, t)$ vs. $y$ for dimensionless times $t = 100, 1200, 1400, 1600$ and $20000$, corre-sponding to alternate frames in Fig. 4. The initial profile (at $t = 100$) is S-rich and the central region stays S-rich even subsequent to axial segregation, i.e., the average order parameter is greater than 0. This is in accordance with most experimental studies [8, 24]. The profiles for dimensionless times $t = 1200, 1400, 1600$ exhibit a growth in the fluctuation amplitude without any coarsening. The final profile (at $t = 20000$) refers to a state in which well-formed bands have already coarsened substantially. Fig. 5(b) shows the temporal evolution of axial profiles for $x = R/2$, which is the approximate location of the interface in the radially-segregated state. The amplitude of fluctuations at early times is comparable to that in Fig. 5(a). Recall that the intrinsic segregation drive is larger for $x = 0$ than $x = R/2$, but this is offset by the off-critical nature of the background for $x = 0$, which suppresses the growth of fluctuations. Finally, Fig. 5(c) shows the axial profiles for $x = R$, corresponding to the drum edge. For the simulation presented here, there is no intrinsic drive to segregate at $x = R$ and the fluctuations at the drum edge are dragged by the fluctuations in the drum centre. Thus, the fluctuations in Fig. 5(c) are of considerably smaller amplitude than those in Figs. 5(a)-(b). Furthermore, the axially segregated state (at $t = 20000$) is still G-rich, as is expected.

We would like to quantify the time-dependence of coarsening in the axial segregation regime. Towards this end, we have computed the average length $L_y(t)$ of S-rich domains at $x = 0, R$. These are obtained from the relevant domain-size distributions which are calculated as an average over 10 independent runs for much larger system sizes than before (i.e., $R = 4$ and $L = 2048$). Fig. 6 shows the time-dependence of the length scales on a log-linear scale. Naively, we may expect that dynamics in the axial segregation regime is effectively 1-dimensional and therefore obeys the 1-dimensional CH equation. It is well-known that domain growth in the 1-dimensional CH equation is logarithmic in time and Fig. 6 shows that the axial length scales for $x = 0, R$ are consistent with logarithmic growth in time. However, we should make two comments here. Firstly, the assumption that domain growth is effectively 1-dimensional is of limited validity, because we know that domain growth at the drum edge is driven by that in the central region. Secondly, a biased plot like Fig. 6 should not be considered as a proof of logarithmic growth!! To conclusively establish a logarithmic growth law, we would need data over at least 3-4 more decades in time and with much better statistics.

## 4  Summary and Discussion

Let us conclude with a summary and discussion of the results in this paper. We have formulated a phenomenological model for the segregation dynamics of binary granular mixtures in a rotating drum. Our model is based on the simple assumptions that (a) the components of the mixture prefer to incline at their natural angles of repose; and (b) the $S$-shaped surface profile is always in local equilibrium, which is determined by the composition of the mixture. These assumptions serve as the basis for a "derivation" of our phenomenological model. We should stress that the "derivation" is of limited validity and should only be understood as a convenient method of arriving at a reasonable coarse-grained model. Of course, it is our underlying expectation that the resultant model is of greater validity than the restrictive "derivation" would suggest.

In this paper, we have presented preliminary analytical results and detailed numerical results for this model. We are able to obtain an analytic approximation for the radially-segregated profile, which works well for parameter regimes of interest to us. Our numerical studies show that the model exhibits a rich dynamical behaviour. In particular, we are able to replicate qualitative experimental observations regarding the dynamics of segregation of granular mixtures. We also make quantitative predictions regarding the functional form of the radially segregated profile; the time-dependence of axial length scales, etc. It would be of considerable relevance if experimentalists could obtain quantitative results for some of these features of segregation.

There are a variety of possible directions for future study. Even at the level of the simple model in Eqs. (21)-(23) studied here, there are a number of outstanding problems. For example, we have still to obtain a comprehensive understanding of the dynamical behaviour of our model in various parameter regimes. A complete solution of the equations governing linear stability of fluctuations about the radially-segregated profile would facilitate an understanding of the "phase diagram" of our simple model. Furthermore, we expect that the incorporation of further generalizations (as discussed in subsection 2.3) will enable an even better replication of experimental results.

The other important problem is to make a quantitative connection between experimental and numerical parameters. At present, we have only established a heuristic relationship between parameters in our model and experimental quantities. We should also stress that the existing experimental results are of a somewhat qualitative nature. We hope that the availability of new experimental techniques (e.g., magnetic resonance imaging) and quantitative predictions from our phenomenological model will motivate experimentalists to obtain more quantitative results.

The next level of modelling should aim for a complete 3-dimensional model, which also accounts for particle motion in the bulk. We have assumed here (as have other authors) that the relevant dynamical behaviour is confined to the 2-dimensional laminar surface layer. This is clearly inadequate in some parameter regimes – for example, at very low rotation frequencies, there is no well-established laminar layer but rather the granular material sloshes about in the drum. Even in the regime where there is a flowing laminar layer, it is not clear that the bulk particle motion merely takes the form of accretion to and depletion from the surface layer. Thus, it is possible that convection rolls under the flowing surface also play a role in segregation [25]. Clearly, the above questions must be answered in the framework of a complete 3-dimensional model. We are presently investigating the form which such a model should take. Again, it would greatly facilitate better modelling if experimentalists could clearly elucidate the nature of particle dynamics in the bulk.

## Acknowledgments

SP is very grateful to H.Hayakawa for his kind invitation to spend a six-month sabbatical in Kyoto, where this work was done. He is also grateful to the Kyoto University Foundation for financial support during his stay in Kyoto. Both authors thank R.Kobayashi for introducing them to this problem and for many useful discussions. They are also grateful to Y.Shiwa and M.Nakagawa for their critical inputs. This work is, in part, supported by a grant from the Ministry of Education, Science, Sports and Culture of Japan (No. 09740314).

# Appendix A

In this appendix, we present a "microscopic" motivation for the expansion presented in subsection 2.2. However, the arguments presented below have a limited range of validity and are best viewed as a guide to good phenomenology.

## A.1 Surface Profile for One-Component Granular Material

For a one-component granular material in a rotating drum, Zik et al. [9] have obtained an expression for the steady-state slope of the surface profile $s(x)$. We will elucidate and expand upon their arguments below. Our discussion in this appendix will also facilitate the physical identification of various expansion parameters introduced in subsection 2.2.

Consider the surface to be defined by its slope $s(x)$ or the angle $\theta(x)$ ($s(x) = \tan[\theta(x)]$) it makes to the horizontal. As stated in subsection 2.1, granular flow occurs in a laminar layer below the surface in the frequency regime of interest to us. We focus on a point $P$ on the surface (with radial coordinate $x$) and consider points vertically below this point in the laminar layer. Let us examine the various forces (per unit area) on the "fluid" at a depth $h$ below $P$. The driving force for motion is gravity, with a component $\rho g h \sin\theta$ parallel to the surface, where $\rho$ is the fluid density (taken to be uniform) and $g$ is the acceleration due to gravity. It should be kept in mind that $\theta$ has an $x$-dependence, though we suppress this for notational convenience. The driving force is opposed by the frictional force $\mu\rho g h \cos\theta$, where $\mu$ is the friction coefficient. If we assume (following Zik et al. [9]) that the granular fluid is Newtonian, then the fluid velocity at the point of interest obeys the following equation :

$$\eta \frac{dv(h)}{dh} = \rho g h (\sin\theta - \mu\cos\theta), \tag{A.1}$$

where $\eta$ is the fluid viscosity. (Of course, the assumption of Newtonian behaviour is of limited validity and we will shortly criticize this assumption.)

Next, we recognize that the fluid velocity goes to 0 at the bottom of the laminar layer, defined by the depth $h_0(x)$. Integrating Eq. (A.1) from $h_0$ to an arbitrary depth $h$, we obtain the velocity profile $v(h)$ as follows :

$$v(h) = \frac{\rho g}{2\eta} \cos\theta (\tan\theta - \mu)(h^2 - h_0^2). \qquad (A.2)$$

The radial mass current (for a unit axial length) flowing below the point P is obtained as follows :

$$\begin{aligned} J_{\rm rad} &= \int_0^{h_0} dh \rho v(h) \\ &= -\frac{\rho^2 g}{3\eta} h_0^3 \cos\theta(\tan\theta - \mu). \end{aligned} \qquad (A.3)$$

The quantity $h_0$ depends upon the angle $\theta(x)$. Zik et al. define the local depth of the laminar layer as corresponding to a point of uniform pressure $p_0$. Thus, $h_0 = p_0/(\rho g \cos\theta)$ and

$$J_{\rm rad} = -\frac{p_0^3}{3\eta\rho g^2}(1 + \tan^2\theta)(\tan\theta - \mu). \qquad (A.4)$$

Next, we consider a simple expression obtained by de Gennes [19] for the surface current of a rotating granular mixture. The argument is quite general and applicable for arbitrary drum-filling fractions. Consider a granular material rotated counterclockwise in a drum as before. Let its surface (assumed to be flat in the present context) be at a vertical distance $d$ from the centre of the drum. The case of half-filling corresponds to $d = 0$. As before, the coordinate $x \in [-\sqrt{R^2 - d^2}, \sqrt{R^2 - d^2}]$ is measured along the surface of the material. At a point $x > 0$, the rotating bulk accretes particles onto the surface layer with a velocity $\omega x$, where we only consider the velocity component normal to the surface. Assuming that all accreted particles join the surface layer, the current input (per unit axial length) is

$$dJ_{\rm rad} = -\rho \omega x dx. \qquad (A.5)$$

The overall radial current at the point $x$ is obtained from mass continuity by integrating the above expression to obtain

$$J_{\text{rad}} = -\frac{\rho\omega}{2}(R^2 - d^2 - x^2). \tag{A.6}$$

The current is maximum at $x = 0$. For points $x < 0$, there is a depletion of surface particles which enter the bulk. A simple adaptation of the above argument shows that Eq. (A.6) is also valid for $x < 0$.

In the steady state, the profile is obtained by equating Eqs. (A.4) and (A.6). Thus, the steady-state profile is obtained from the condition

$$\frac{p_0^3}{3\eta\rho g^2}(1 + \tan^2\theta)(\tan\theta - \mu) = \frac{\rho\omega}{2}(R^2 - d^2 - x^2). \tag{A.7}$$

Before we proceed, we would like to discuss the validity of the above approach of Zik et al. [9]. Apart from missing factors of the mass density $\rho$, which we have attempted to rectify in the above discussion, their argument is also open to more serious criticism. Some of the defects of this argument are as follows :

(a) As we have mentioned earlier, Eq. (A.1) assumes that the fluid is Newtonian, whereas the true behaviour of a granular fluid is generally non-Newtonian [1]. Thus, in general, Eq. (A.1) should be replaced by its non-Newtonian equivalent

$$\eta \frac{dv(h)}{dh}\left|\frac{dv(h)}{dh}\right| = \rho g h(\sin\theta - \mu\cos\theta). \tag{A.8}$$

(b) The profile resulting from Eq. (A.7) is too symmetric and has the same slopes for points labelled by $x$ and $-x$. However, in experimental situations, the $S$-shaped profile is steeper at the top ($x \sim \sqrt{R^2 - d^2}$) than at the bottom ($x \sim -\sqrt{R^2 - d^2}$). Zik et al. [9] argue that this discrepancy can be rectified by accounting for the difference between the static and dynamic friction coefficients, which are relevant at the top and bottom of the profile, respectively.

Thus, it should be clear that the profile obtained from Eq. (A.7) is not microscopically correct. Nevertheless, the above arguments can serve as a guide to good phenomenology and that is the sense in which we shall use them in this paper.

It is straightforward to obtain a solution for Eq. (A.7) by solving the cubic equation [20] to determine the local slope $s(x) = \tan[\theta(x)]$. The cubic equation has only one real solution, which is

$$\begin{aligned}
s(x) &= \frac{\mu}{3} + \left\{ \frac{\mu}{3} + \frac{\mu^3}{27} + \frac{A}{2} + \left[ \frac{(1+\mu^2)^2}{27} + \frac{A^2}{4} + \frac{A}{3}\mu + \frac{A}{27}\mu^3 \right]^{1/2} \right\}^{1/3} \\
&+ \left\{ \frac{\mu}{3} + \frac{\mu^3}{27} + \frac{A}{2} - \left[ \frac{(1+\mu^2)^2}{27} + \frac{A^2}{4} + \frac{A}{3}\mu + \frac{A}{27}\mu^3 \right]^{1/2} \right\}^{1/3}, \quad \text{(A.9)}
\end{aligned}$$

where we have set $d = 0$ (corresponding to half-filling) and introduced the notation

$$A = \frac{3\eta\rho^2 g^2 \omega}{2p_0^3}(R^2 - x^2) \equiv \bar{A}\eta. \quad \text{(A.10)}$$

## A.2 Generalization to Two-Component Granular Mixtures

We are interested in the surface profile for a two-component granular mixture. As stated in subsection 2.1, we will assume that the profile is always in local equilibrium. Thus, at a point with composition $\psi$, the slope $s(\vec{r}, t)$ is determined by Eq. (A.9) with the simplest appropriate generalizations of $\eta$ and $\mu$, viz.,

$$\begin{aligned}
\mu &= \frac{\mu_S + \mu_G}{2} + \frac{\mu_S - \mu_G}{2}\psi \equiv \mu_+ + \mu_-\psi, \\
\eta &= \frac{\eta_S + \eta_G}{2} + \frac{\eta_S - \eta_G}{2}\psi \equiv \eta_+ + \eta_-\psi. \quad \text{(A.11)}
\end{aligned}$$

It should be kept in mind that the axial and time-dependence of $s(\vec{r}, t)$ are entirely a consequence of the corresponding dependence of the order parameter $\psi(\vec{r}, t)$. Furthermore, in the absence of rotation ($A = 0$), Eq. (A.9) trivially reduces to the expected solution $s(\vec{r}, t) = \mu$.

Recall that our reference frame in Fig. 1 is inclined to the horizontal at an angle $\theta_1$, which is the average repose angle of the rotated homogeneous mixture. This angle can be determined from Eq. (A.9) as

$$\tan\theta_1 = \frac{1}{2R}\int_{-R}^{R} dx s_{\text{hom}}(x), \quad (A.12)$$

where $s_{\text{hom}}(x)$ is the form of $s(\vec{r},t)$ for the case of uniform $\mu$ and $\eta$.

In the context of the solution $s(\vec{r},t)$ obtained from Eq. (A.9), it is simple to demonstrate that $s(\vec{r},t) - s_G(x) \geq 0$, as we expect physically. Recall that $s(\vec{r},t)$ satisfies Eq. (A.7), viz.,

$$(1+s^2)(s-\mu) = \bar{A}\eta, \quad (A.13)$$

where we suppress the $(\vec{r},t)$-dependence of various quantities. It is straightforward to calculate $ds/d\psi$ from Eq. (A.13) as follows :

$$\frac{ds}{d\psi} = \frac{\bar{A}\eta_- + (1+s^2)\mu_-}{1+s^2+\frac{2\bar{A}\eta s}{(1+s^2)}} > 0, \quad (A.14)$$

where we have used Eq. (A.11). Thus, $s(\psi)$ increases monotonically with $\psi$ and we have $s(\vec{r},t) - s_G(x) \geq 0$.

It is convenient to obtain a more tractable form for $s(\vec{r},t)$ by expanding in "small" quantities. This enables a physical identification of various parameters introduced in subsection 2.2.

First, we expand the expression for $s(\vec{r},t)$ about $A = 0$ (or $\omega = 0$), corresponding to the case with no rotation. The resultant expansion is

$$s(\vec{r},t) = \mu + \frac{1}{1+\mu^2}A - \frac{2\mu}{(1+\mu^2)^3}A^2 + O(A^3). \quad (A.15)$$

We will subsequently retain only terms linear in $A$. We should caution the reader that our 2-dimensional model does not contain the correct experimental behaviour at very low rotation frequencies. Experimentally, one obtains a "sloshing" about of the granular material in this regime [18]. This behaviour has obviously been eliminated in our assumption of a steady-state surface profile. We believe that this oscillatory behaviour can only be recovered in the complete 3-dimensional model supplemented with the correct boundary conditions on the drum and, more importantly, the free surface. In the present context, though we have expanded about $A = 0$, the experimentally relevant regime in our model corresponds to intermediate rotation frequencies, where segregation takes place. However, we should emphasize that this is consistent in the context of our model, which assumes the existence of a laminar flow at all rotation frequencies.

Next, we expand the expression for $s(\vec{r}, t)$ in Eq. (A.15) in the weak-segregation limit, corresponding to small values of the order parameter. This yields the following expansion :

$$s(\vec{r},t) = \mu_+ + \frac{\bar{A}\eta_+}{1+\mu_+^2} + \left[\mu_- + \frac{\bar{A}}{1+\mu_+^2}\left(\eta_- - \frac{2\mu_+\eta_+}{1+\mu_+^2}\mu_-\right)\right]\psi(\vec{r},t) + O(\psi^2). \tag{A.16}$$

A comparison of Eq. (A.16) with Eq. (10) facilitates identification of the various parameters introduced in subsection 2.2. In this paper, we consider a situation in which the relative viscosity difference between S and G is appreciably large. This is physically reasonable in the context of sand ($\mu_S \simeq 0.73$) and glass ($\mu_G \simeq 0.58$), though we are not aware of precise evaluations of viscosity for these granular fluids. In such a situation, the quantity $\gamma = \eta_- - 2\mu_+\eta_+\mu_-/(1+\mu_+^2) > 0$. However, this is not essential and the case where $\gamma < 0$ is also of possible experimental relevance.

# Figure Captions

Fig. 1 : Schematic of the radial cross-section of a drum half-filled (shaded region) with a mixture of rough (sand or S) and smooth (glass or G) granular materials. The drum has a radius $R$ and is rotated counterclockwise with angular velocity $\omega$ about its horizontal axis. The surface profile of the granular mixture has a characteristic $S$-shaped profile. The 2-dimensional reference frame is chosen so that the $x$-axis (radial direction) points along the average of the $S$-shaped profile for a homogeneous rotated mixture. The $y$-axis points along the axis of the drum into the plane of the paper.

Fig. 2 : Temporal evolution of the radial order parameter profile in a 1-dimensional simulation of our model in Eqs. (21)-(23). The parameter values were $b = 0.9$ and $a = 0.03$; and the drum radius was $R = 4$ in dimensionless units. Other details of the simulation are provided in the text. We plot $\psi(x,t)$ vs. $x$ for dimensionless times $t = 0.5, 1, 2, 4$ and 10, denoted by the indicated symbols. The solid line denotes the analytic approximation described in the text for the static radial profile.

Fig. 3 : Static radial profiles $\psi_s(x)$ vs. $x$, obtained from numerical simulations of the 1-dimensional version of Eqs. (21)-(23). We present results for parameter values $b = 0.9$ and $a = 0.01, 0.03, 0.05$ and $0.1$, denoted by the indicated symbols. The corresponding analytic approximation for the static profile is superposed as a solid line on the data sets for $a = 0.01, 0.03, 0.05$.

Fig. 4 : Temporal evolution of the order parameter from a 2-dimensional simulation of our model in Eqs. (21)-(23). The parameter values were $b = 0.9$ and $a = 0.03$. Other simulation details are provided in the text. The initial condition for the run consisted of an order parameter field which was uniformly and randomly distributed (with amplitude 0.05)

about a zero background. This initial condition mimics a homogeneous mixture with equal amounts of S and G. In the frames shown, regions rich in S ($\psi > 0$) are marked in black and regions rich in G ($\psi < 0$) are not marked. The frames are shown at dimensionless times $t = 100$, showing the radially-segregated structure; $t =1100$-$1500$, showing the destabilization of the radially-segregated state; and $t = 1600, 10000$ and $20000$, showing the coarsening of the axial bands.

Fig. 5 : Axial variation of the order parameter profiles for the evolution depicted in Fig. 4. We plot $\psi(x, y, t)$ vs. $y$ for dimensionless times $t = 100, 1200, 1400, 1600$ and $20000$, corresponding to alternate frames in Fig. 4. The values of $x$ are (a) $x = 0$, at the drum centre; (b) $x = R/2$, approximately at the interfacial position for the radially segregated profile; and (c) $x = R$, at the edge of the drum.

Fig. 6 : Time-dependence of the axial length scales $L_y(t)$ for the S-rich regions. These length scales are computed as averages of the relevant domain-size distribution functions, calculated as described in the text. We plot $L_y(t)$ vs. $t$ on a log-linear scale for $x = 0$ and $x = R$, denoted by the indicated symbols.

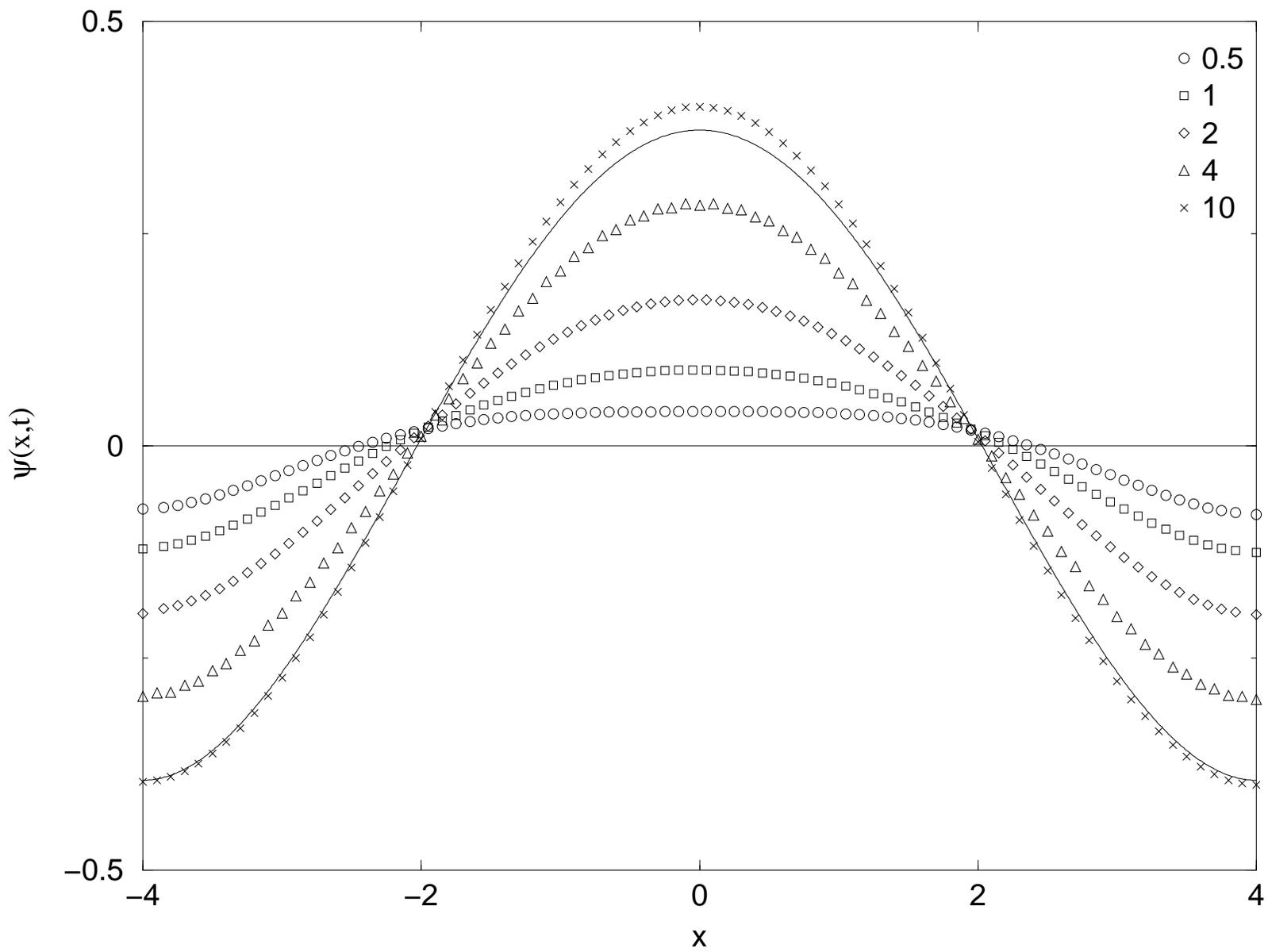

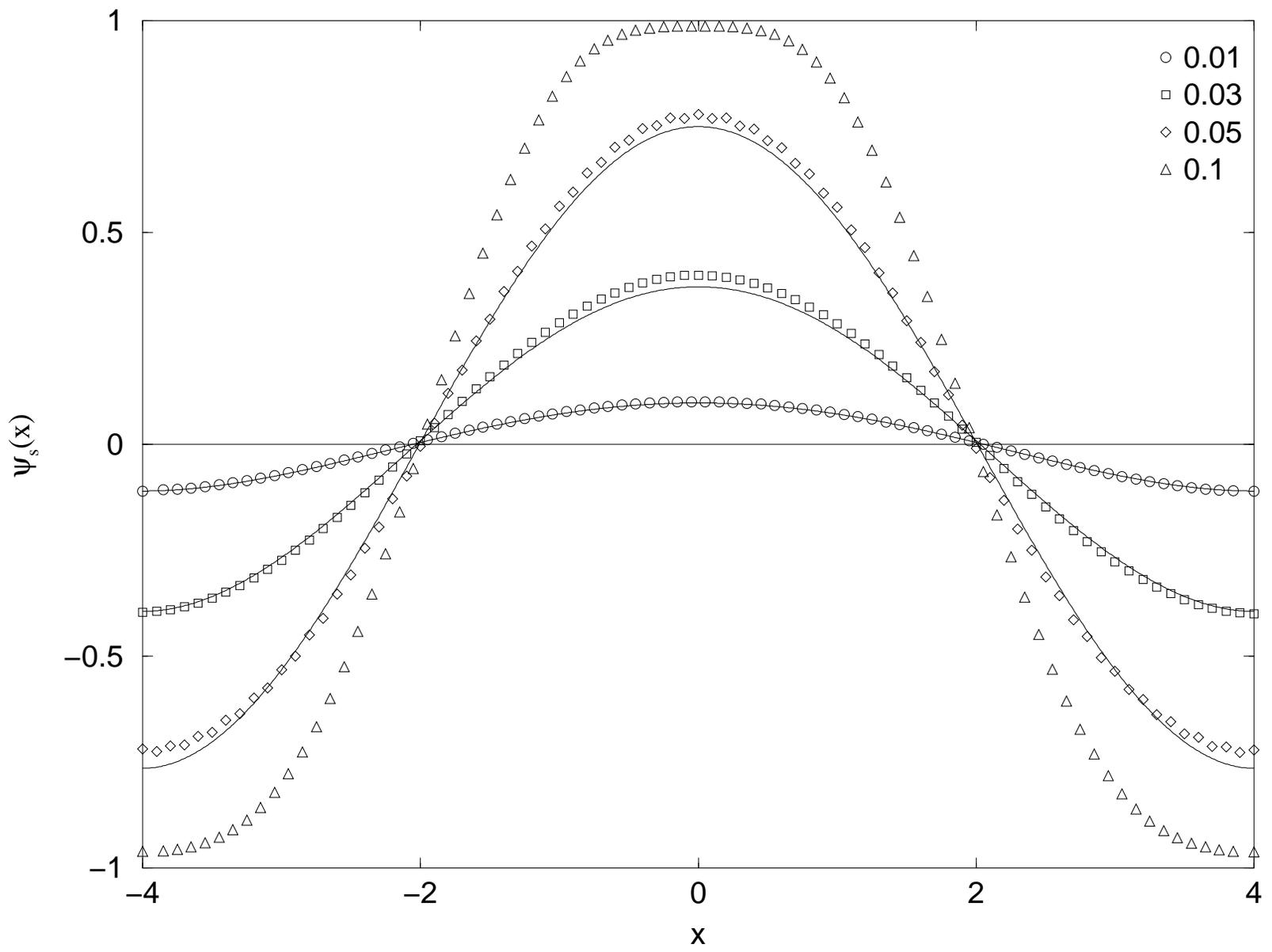

| 100 | 1100 | 1200 | 1300 | 1400 | 1500 | 1600 | 10000 | 20000 |
|---|---|---|---|---|---|---|---|---|
| 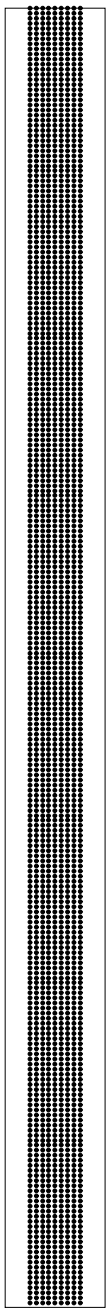 | 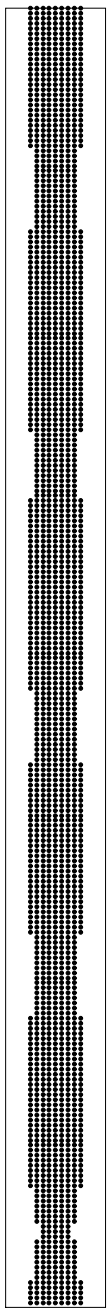 | 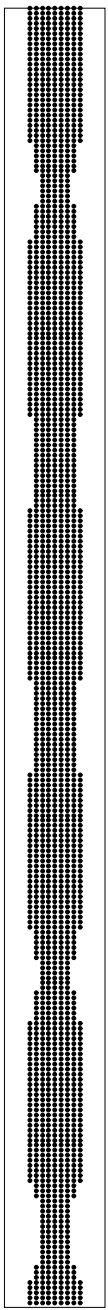 | 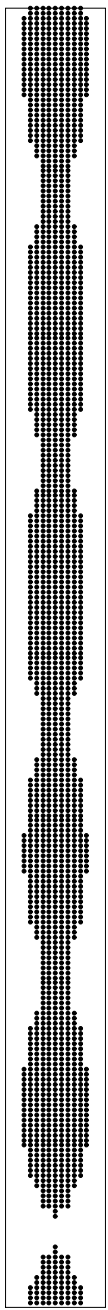 | 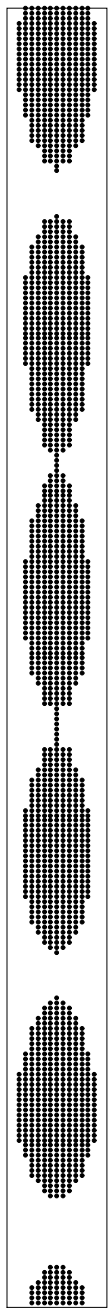 | 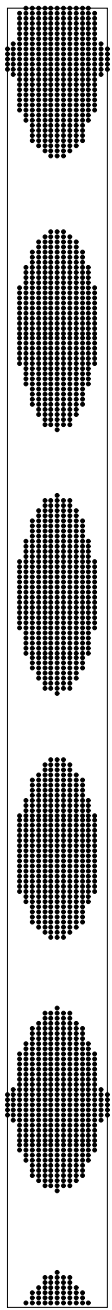 | 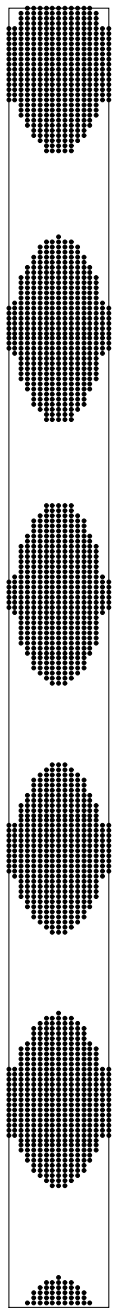 | 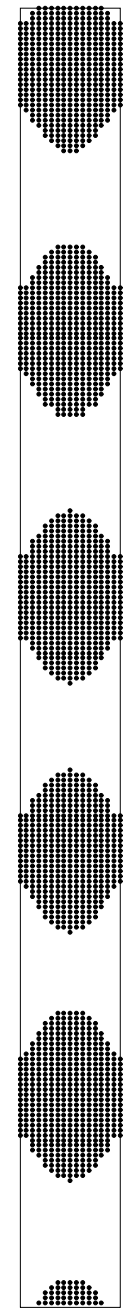 | 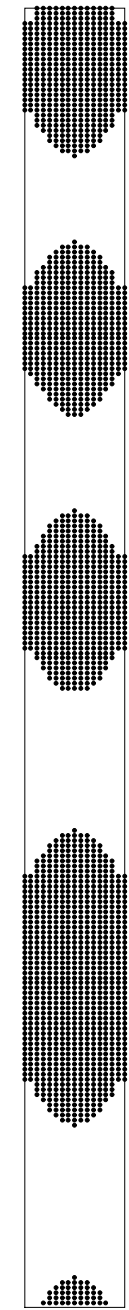 |

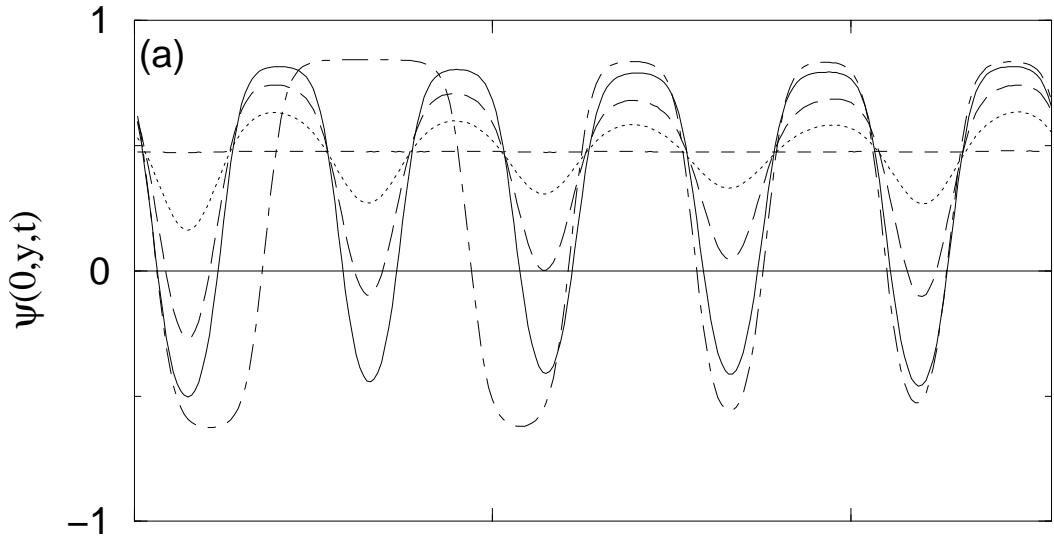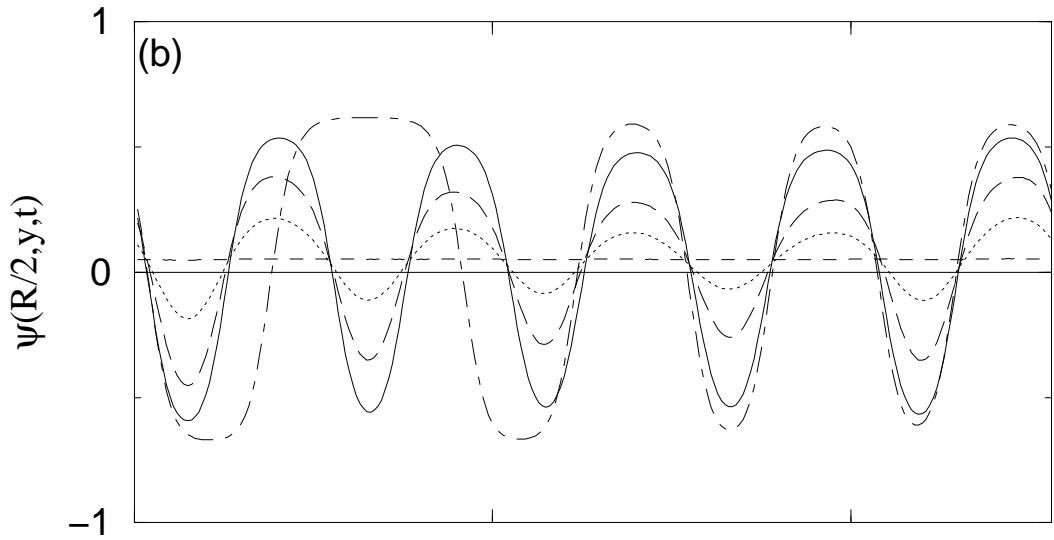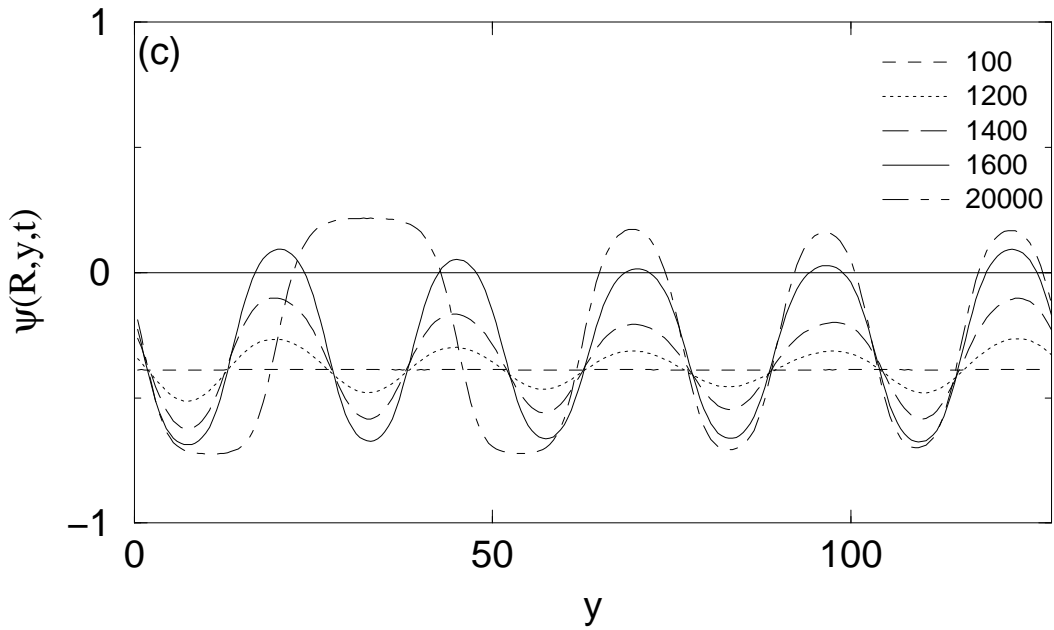

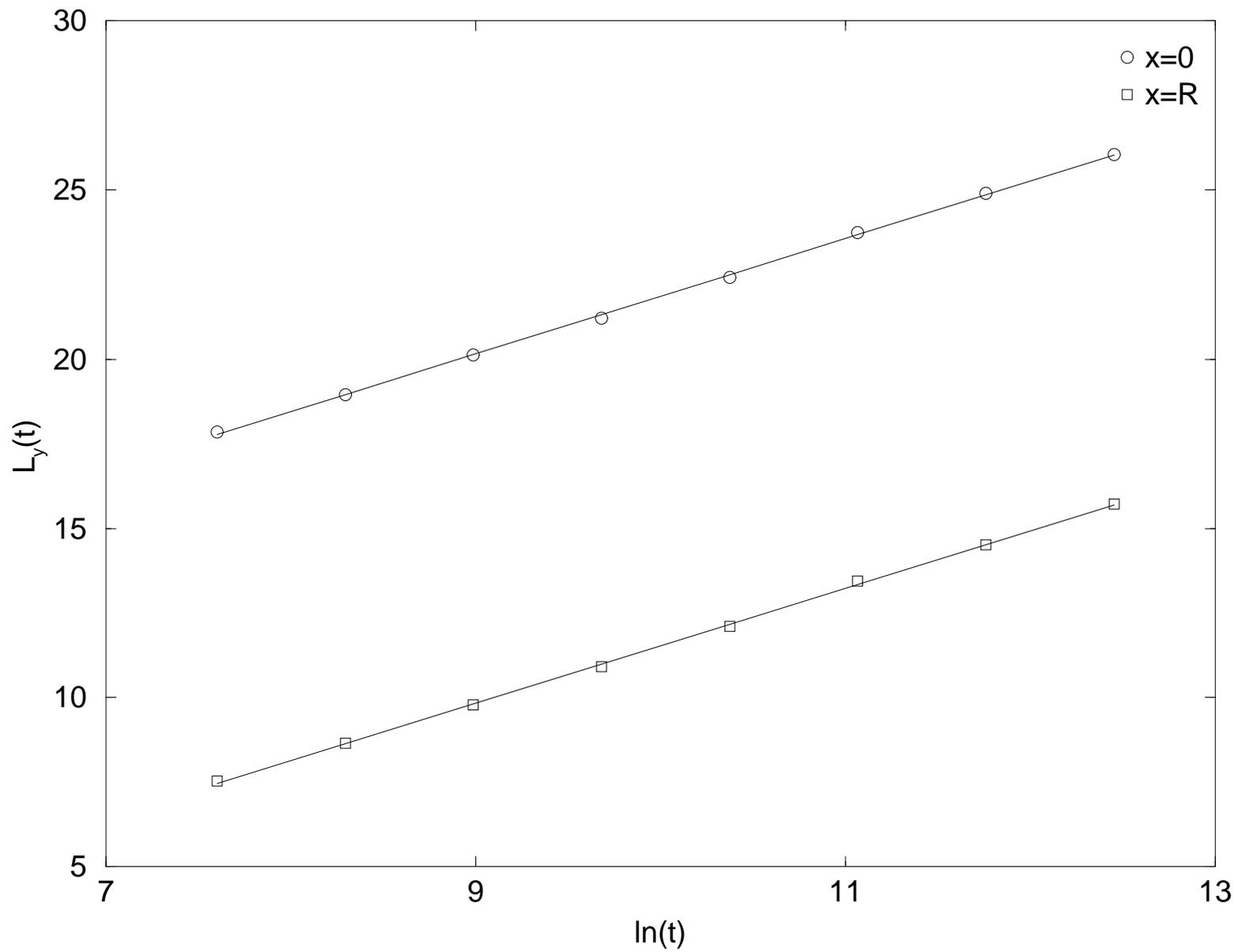